\documentclass[12pt,a4paper]{article}
\usepackage{parskip}
\usepackage{latexsym}
\usepackage{times}
\usepackage{color}
\usepackage{graphicx}
\usepackage{amsmath}

\begin{document}
\title{Growth-induced mass flows \\ in fungal networks.}
\author{Luke L.M. Heaton$^{1,2}$, Eduardo L\'{o}pez$^{2,3}$, \\
Philip K. Maini$^{3,4,5}$, Mark D. Fricker$^{3,6}$, Nick S. Jones$^{2,3,5}$}
\maketitle

{\footnotesize $^{1}$ LSI DTC, Wolfson Building, University of Oxford, Parks Road, Oxford, OX1 3QD, UK \\
$^{2}$ Physics Department, Clarendon Laboratory, University of Oxford, Parks Road, Oxford, OX1 3PU, UK \\
$^{3}$ CABDyN Complexity Centre, Sa\"{\i}d Business School, University of Oxford, Park End Street, Oxford, OX1 1HP, UK \\
$^{4}$ Centre for Mathematical Biology, Mathematical Institute, University of Oxford, 24-29 St Giles', Oxford, OX1 3LB, UK \\
$^{5}$ Oxford Centre for Integrative Systems Biology, Department of Biochemistry, University of Oxford, South Parks Road, Oxford, OX1 3QU, UK \\
$^{6}$ Department of Plant Sciences, University of Oxford, South Parks Road, Oxford, OX1 3RB, UK}

\section*{Abstract}


Cord-forming fungi form extensive networks that continuously adapt to maintain an efficient transport system. As osmotically driven water uptake is often distal from the tips, and aqueous fluids are incompressible, we propose that growth induces mass flows across the mycelium, whether or not there are intrahyphal concentration gradients. We imaged the temporal evolution of networks formed by \emph{Phanerochaete velutina}, and at each stage calculated the unique set of currents that account for the observed changes in cord volume, while minimising the work required to overcome viscous drag. Predicted speeds were in reasonable agreement with experimental data, and the pressure gradients needed to produce these flows are small. Furthermore, cords that were predicted to carry fast-moving or large currents were significantly more likely to increase in size than cords with slow-moving or small currents. The incompressibility of the fluids within fungi means there is a rapid global response to local fluid movements. Hence velocity of fluid flow is a local signal that conveys quasi-global information about the role of a cord within the mycelium. We suggest that fluid incompressibility and the coupling of growth and mass flow are critical physical features that enable the development of efficient, adaptive, biological transport networks.

\textbf{Key words:} Mycelial modelling; nutrient translocation; complex networks..

\section*{Introduction}

Multi-cellular organisms have evolved sophisticated systems to supply individual cells with the resources necessary for survival. Plants circulate nutrients through the xylem and phloem, driving mass flows in the xylem by transpiration from the leaves. They also actively maintain osmotic gradients along the phloem, inducing a flow of sap from sources, where water is drawn from the surrounding tissue into the sieve-tubes of the phloem, to sinks, where water leaves the phloem (Nelson, 2003; Nobel, 1991). Animals utilise hearts or contractile regions to circulate blood through hierarchical, fractal-like vascular systems (Savage \emph{et al}. 2008; Sherman 1981). In contrast, transport through fungal mycelial networks is poorly understood.

Foraging fungal mycelia continuously re-model their morphology, and as a transport network the mycelium must adapt to changing local environmental conditions and patchy resource availability (Bebber \emph{et al}. 2007; Boddy 1999; Cairney 1992; Fricker 2007; Jennings 1987; Falconer \emph{et al}. 2005, 2007; Gow \& Gadd 1995). Hyphae grow by tip extension and then branch sub-apically to form a diffuse tree-like mycelium (Gooday 1995; Money 1997, 2008; Steinberg 2006). As the colony continues to grow, hyphal fusions or anastomoses occur, producing a more reticulate, net-like structure. In cord-forming fungi, hyphal aggregates subsequently develop, and undergo limited differentiation to yield specialized high-conductivity channels that are often well-insulated from the environment (Cairney 1992; Jennings 1987; Rayner \emph{et al}. 1991). These `cords' may thicken or thin over time, and they contain numerous rigid, hollow vessel hyphae with very few septal pores (Eamus 1985). Other regions of the mycelium regress, probably by autophagy, to recycle redundant material to support new growth (Falconer 2005, 2007; Fricker 2007; Olsson 2001).

Whilst direct uptake and intra-hyphal nutrient diffusion may be sufficient to sustain short-range local growth when resources are abundant (Olsson 2001), long-distance translocation is required to deliver nutrients at a sufficient rate to growing tips, particularly in non-resource restricted fungi that are too large to distribute nutrients through diffusion alone (Boswell \emph{et al}. 2002, 2003a, 2003b, 2007; Clipson \emph{et al}., 1987; Eamus \& Jennings, 1984; Wells \emph{et al}. 1995, Wells \& Boddy 1995). Remarkably little is known about the mechanism(s) underpinning such long-distance nutrient translocation, or the quantitative contribution of different potential transport pathways, such as cytoplasmic streaming, vesicle transport or mass flow, to net fluxes and overall nutrient dynamics (Cairney 1992; Fricker 2007; Jennings 1987).

\begin{figure}[htbp]
\begin{center}
\includegraphics[width=13.2cm]{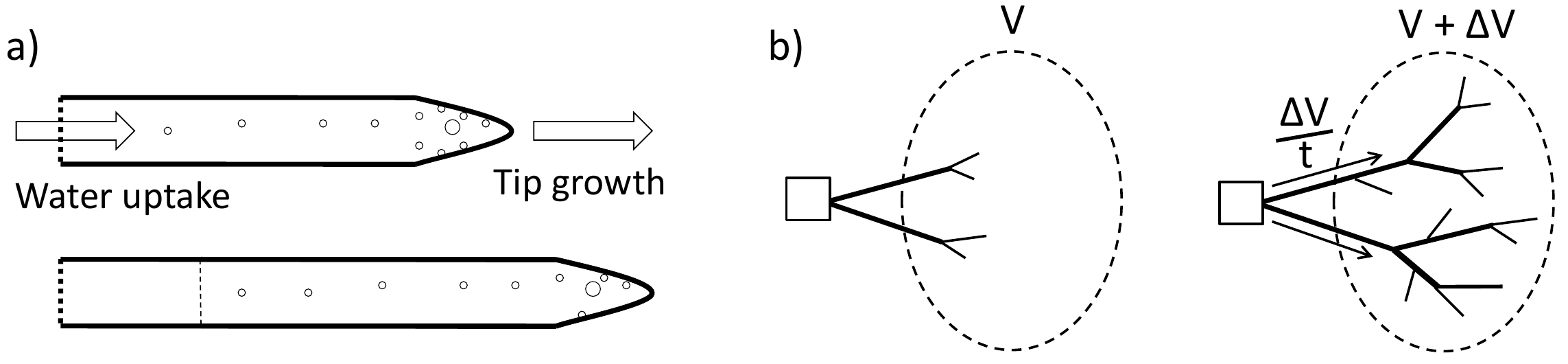}
\caption{\textbf{Physical principles of growth-induced mass flow.} \newline
\textbf{a)} Turgor pressure is induced by an osmotic gradient at the site of water uptake. Vesicles (circles) move towards the tip faster than it recedes, while the cytosol behind the growing tip moves forward at the rate of tip growth (Lew, 2005). The conservation of volume dictates that as the tip expands, fluid flows towards the tips from the site of water uptake. This mass flow demonstrates the presence of a pressure gradient. \newline \newline
\textbf{b)} Suppose that a fungus grows out of an inoculum (square) and into a region (oval). Some of the material that becomes part of the fungi may come from within the oval region. The rest of the material must have travelled along the cords (edges) that cross the region's boundary. If the volume of fungi within the region increases by $\Delta V$ over a period of time $t$, and none of the material is drawn from within the region, it follows that the net current flowing into the region is $\frac{\Delta V}{t}$. Furthermore, if the total cross-sectional area of the boundary crossing cords is $a$, the mean velocity of flow will be $\frac{\Delta V}{at}$.}
\label{growth_diagrams_1}
\end{center}
\end{figure}

We suggest that regardless of intra-hyphal concentration gradients, mass flow only takes place when water is able to exit the translocation pathway through either localised exudation (e.g. \emph{Serpula lacrymans}), evaporation, or by moving into a region of new growth. In this paper we quantify the last of these phenomena, which we have termed growth-induced mass flow. By way of physical analogy consider a rigid tube filled with salty water that is blocked at one end by a semi-permeable membrane, while the other is blocked by a thin rubber cap. If this apparatus is submerged in water, the osmotic gradient across the semi-permeable membrane will induce turgor pressure, and the pressure within the tube will force the rubber cap to bulge outwards.  As aqueous fluids are essentially incompressible,  the column of fluid within the tube can only move forward at the same rate as the rubber cap, and this movement indicates the presence of a pressure gradient (Fig. \ref{growth_diagrams_1}a).

Injected oil droplets in individual hyphae of \emph{Neurospora} provide evidence for this kind of growth-coupled mass flow, as the average rate of movement ($\sim0.5 \mu \textrm{ms}^{-1}$) matches the  rate of tip extension (Lew, 2005). Such movement would be consistent with mass flow driven by the continuous sub-apical water influx required to sustain volume increases at the tip during growth. Taken together, these lines of evidence suggest growth-coupled mass flow may have a significant role in water and nutrient translocation in larger mycelial systems.

To quantify the scale of growth-induced mass flow we developed two models, the `uniform model' and the `time-lapse model', and applied these models to measured examples of mycelial growth. To obtain a sample of fungal networks we allowed \emph{P. velutina} to grow over experimental microcosms for a four week period, taking photographs every three days. An image analysis program was then used to convert the sequence of photographs into a sequence of networks, comprised of cords of measured length and volume. Given  the measured volumes and changes in volume, we used these models to calculate a current for each edge. The currents calculated by the uniform model reflect the topology of the network. The time-lapse model produces an estimate for the minimum flow of material that is consistent with the measured changes in volume, under the assumption that the inoculum is the sole source of water and nutrients.

We found that the cords with higher currents or higher speeds of mass flow were more likely to increase in thickness than the other cords (see Fig. \ref{speed_and_current_v_change}). This suggests that our model is correctly identifying the high current cords, since thickening the high current cords is an efficient way to remodel a fungal network. This follows because increasing the thickness and conductance of any cord will reduce the cost of overcoming viscous drag, but it is much more efficient to thicken cords which carry large currents rather than thickening the cords with small currents.

\section*{Model development}

By definition a network is a collection of nodes together with the edges that connect those nodes. A node that is connected to a single edge is called a `tip', and we refer to all other edges as `cords'. By this definition a cord is a cylindrical, linear structure, while any branching fungal form is described as a network of cords. We use letters such as $i$ and $j$ to index nodes, and pairs of letters $ij$ index the edge (or cord) between nodes $i$ and $j$. In this paper we describe two models that can be unfolded over a given network: the `uniform model' and the `time-lapse model'. Details are supplied in later sections, and a preview can be gained by a glance at Fig. 2 or Fig. S1. Both models take as their inputs experimentally determined networks extracted from images; how this image data is used to calculate currents differs between the two cases. Furthermore, both models have been defined for water uptake at a single location (the inoculum), but could be adapted for growth involving more than one source of nutrients and water. 

The uniform model uses only a small amount of information from a single imaged network: it throws away all detail about the thickness of cords (but not their lengths) and supposes they are of uniform thickness. It calculates currents by supposing that all tips grow at the same rate. Given any network, the output of the uniform model is a `current' for each cord. This quantity reflects the topological location of the cords, and the currents predicted by the uniform model reflect the number of tips `downstream' from each of the given cords (see Supplementary Information Section 1, Fig. S1a). We found that under the uniform model, cords with a large current tend to be thicker than cords with a small current.

While the uniform model calculates currents by assuming that all tips grow at the same rate, the time-lapse model uses observed changes in volume between successive pairs of aligned networks. We effectively derive a minimal set of currents that are consistent with the measured growth. A key principle behind the time-lapse model is  that if an object (e.g. a thickening cord) is composed of incompressible material, the rate of increase in the volume of that object must equal the rate of flow into that object minus the rate of flow out of that object. Growth requires the flow of materials, and the time-lapse model was designed to quantify the extent to which changes in volume generate mass flows through the supporting mycelial network. The currents we will calculate represent a minimal total flux, found by calculating the unique set of currents that account for the observed changes in cord volume, while minimising the work required to overcome viscous drag.

\vspace{1.5cm}

\paragraph{Pressure gradients, hydraulic conductance, current and velocity}
\quad \newline
The relationships between pressure gradients, hydraulic conductance, current and velocity
are fundamental to understanding fluid flows in plants and fungi. We note, however, that there is no close analogy between flows in open-ended plant vessels which are permeable along their lengths and close-ended fungal vessels with hydrophobic coatings. In plants, concentration gradients draw water from the surrounding tissue into, and out of, the xylem and phloem all along their lengths. By contrast, fungal cords have hydrophobic coatings, and we suppose that such cords do not directly draw fluid from their surrounding environment (in our model, water uptake only occurs at the inoculum). In plants one expects flow between two points to be explained by a chemical potential difference. Similarly, in fungi there is a chemical potential difference between the environment and the hyphae responsible for water and nutrient uptake, and the fungus must do work to maintain such osmotic gradients (Amir \emph{et al}. 1995a, 1995b; Bancal \& Soltani 2002; Lew \emph{et al}. 2004). However, flows between two points \emph{within} the fungus need not be caused by chemical potential differences between those points. The physical effects of growth, turgor pressure and fluid incompressibility may suffice to create flows across the mycelium. 

'Our models require no commitments concerning the mechanisms of fungal growth: fluid incompressibility means that volume is conserved regardless of the mechanisms driving fungal fluid flows. We make the simplifying assumption that the inoculum is the sole source of extra fluid, and the presence of this fluid enables the volume change we observe in the growing fungus. Of the possible set of current flows consistent with the observed changes in volume, we parsimoniously identify the unique flows which minimize energy losses to resistance. Since the fluid flows within fungi are laminar (see SI Section 3), it is appropriate to apply the Hagen-Poiseuille equation, which accurately
describes the laminar flow of incompressible fluids along an insulated tube.

By definition (and by analogy with Ohm's law) the current in a cord must be equal to the pressure drop times the conductance (Eamus 1985; Nobel 1991). The size of the conducting vessels within a cord does not vary significantly with the size of the cord. We are therefore motivated to assume that all cords contain tubes of some fixed radius. We also assume that the number of tubes per unit of cross-sectional area is constant throughout the network. The Hagen-Poiseuille equation tells us that the total current $f$ through a cord comprised of $n$ tubes will satisfy the equation
\begin{equation}
f =  \Delta P C =  \Delta P \frac{n \pi r^{4}}{8 \eta l} = \sigma \Delta P A,
\label{current_pressure_conductance}
\end{equation}
where $\Delta P$ is the pressure drop between the ends of the cord,  $C$ is the hydraulic conductance of cord, $r$ is the radius of the tubes, $\eta$ is the dynamic viscosity of the fluid, $l$ is the length of the cord, $A$ is the cross-sectional area of the cord and $\sigma$ is a constant of proportionality. The conductance of a cord is proportional to the number of tubes it contains,  and if we assume a constant density of tubes, conductance is proportional to cross-sectional area.

\paragraph{Modelling fluid flows: the uniform model}
\quad \newline
The `uniform model' described in this section takes as its input an empirically observed fungal network, where each cylindrical cord in the network has a measured length. We have called it the uniform model because we assume uniform conductance (cross-sectional area) throughout the network, and assume a unit current outflow at every tip.

The rules governing the uniform model are as follows:

\begin{enumerate}
\item We assume unit growth at every tip, so there is a unit current flowing towards each tip.
\item The net current flowing away from the inoculum is equal to the total number of tips. In other words, water uptake occurs at the inoculum, and the rate of water uptake equals the total rate of growth.
\item All cords have the same resistance per unit length. In other words, the conductance of each cord is inversely proportional to its length.
\end{enumerate}
	
	Current effectively enters the network at the inoculum (source) and exits at the tips (sinks). Elsewhere the currents of an incompressible fluid must obey Kirchhoff's law, which states that the total current flowing into a point must equal the total current flowing out. In other words, the net current must be zero. It follows that where $q_{i}$ is the net current flowing out of node $i$,
\begin{equation}
q_{i} = \Bigg\{	
\begin{array}{cl}
-1 & \textrm{ if node $i$ is a tip} \\
m & \textrm{ if node $i$ is the inoculum (where $m$ is the number of tips)} \\
0 & \textrm{ otherwise.}
\end{array}
\end{equation}

Note that the net current at the inoculum is positive, because the flow is directed away from the inoculum. Equation (\ref{current_pressure_conductance}) tells us that the current in a cord is equal to the pressure drop times the conductance. We can use this fact to sum the currents that flow in or out of node $i$, so we have
\begin{equation}
\sum_{j} ( p_{i} - p_{j} ) C_{ij} = q_{i},
\label{p_C_and_q}
\end{equation}
where $p_{i}$ is the pressure at node $i$ and $C_{ij}$ is the conductance of the cord between nodes $i$ and $j$. In the uniform model,
\begin{equation}
C_{ij}  = \Bigg\{	
\begin{array}{cl}
0 & \textrm{ if nodes $i$ and $j$ are not directly connected, and } \\
1/l_{ij} & \textrm{ if there is a cord $ij$ of length $l_{ij}$.}
\end{array}
\end{equation}
Given the conductance of each cord and the net current flowing out of each node, we can
uniquely determine the pressure difference between any pair of nodes (see SI Section 2, or Grimmett \& Kesten 1984; Lopez \emph{et al}. 2005). Furthermore, by Equation (\ref{current_pressure_conductance}) we can uniquely determine the currents
in the network.

\paragraph{Modelling currents induced by changes in volume: the time-lapse model}
\quad \newline
Each experiment yielded a sequence of eleven digitised networks. Unlike the uniform model, the time-lapse model does not assume that the tips are growing at constant rate. Instead we calculate currents by looking at how each network changes in the next time-step. The networks in each sequence must be aligned, and all nodes are considered to be present at all times (so some nodes in a network may not be connected to any edges/cords). We know the time-lapse $t$ between the earlier and later networks, and each cord has a measured length $l_{ij}$, a volume in the earlier network $u_{ij}$ (though this volume may equal zero), a cross-sectional area in the earlier network $a_{ij}=u_{ij}/l_{ij}$, and a volume in the later network $v_{ij}$.

To calculate the currents we must know the relative conductances of the cords, and the net current at each node (note that since edges can thicken or narrow they can become sinks and sources like the tips and inoculum). By Equation (1) the conductance of cords is proportional to their cross-sectional area. Where $\sigma$ is an arbitrary constant of proportionality and $\delta$ is small compared to the cross-sectional areas of the cords, the conductance $C_{ij}$ of cord $ij$ is defined to be
\begin{equation}
C_{ij} = \Bigg\{
\begin{array}{ll}
0 & \textrm{if nodes $i$ and $j$ are not connected,} \\
\sigma a_{ij}/l_{ij}  & \textrm{if $i$ and $j$ are connected in the earlier network, and} \\
\sigma \delta/l_{ij}  & \textrm{if $i$ and $j$ are only connected in the later network.}
\end{array}
\label{C_def}
\end{equation}

In the uniform model the inoculum is the source, and each tip is a sink. As noted, in the case where the volume of cords changes over time, thickening cords are sinks, while thinning cords and the inoculum are sources. Now, the volume of cord $ij$ changes from $u_{ij}$ to $v_{ij}$ over a period of time $t$. Therefore the current flowing into $ij$ must be $(v_{ij} - u_{ij})/t$ greater than the current flowing out of $ij$. As a simplifying assumption, we put half of edge $ij$'s demand for current (sink) at node $i$, and half at node $j$. In other words, we suppose that the current flowing into node $i$ is $(v_{ij} - u_{ij})/2t$ greater than the current flowing out of node $i$, and likewise for node $j$ (see Equation \ref{q_def}).

In both the uniform model and the time-lapse model the conservation of volume leads us to suppose that the rate of water uptake equals the total rate of growth. Hence the net current at the inoculum must be such that the total net current is zero. There is one final consideration behind the definition of the net currents in the time-lapse model.
To make an unbiased analysis of the relationship between current and changes in cross-sectional area, we calculate the current induced in cord $\alpha\beta$ by the changes in volumes of the all the cords excluding cord $\alpha\beta$ itself. For these reasons, when we are calculating the growth-induced current through the cord $\alpha\beta$, the net current flowing out of node $i$ is defined to be

\begin{equation}
q_{i} = \Bigg\{
\begin{array}{ll}
-\sum_{j \neq i} q_{j} & \textrm{if node $i$ is the inoculum, and} \\
& \\
\sum_{ij \neq \alpha\beta} \frac{u_{ij}-v_{ij}}{2t}  & \textrm{otherwise.} \\
\end{array}
\label{q_def}
\end{equation}

Note that the first sum is over the set of all nodes, while the second sum is over the set of all the cords $ij$ directly connected to node $i$. As in the uniform model, we can use the conductance of each cord and the net current flowing out of each node to uniquely determine the pressure difference between any pair of nodes (see SI Section 2 for details on this calculation, and a discussion of the model parameters). Given the pressure drop between the nodes, we can use Equation (\ref{current_pressure_conductance}) to uniquely determine the current in each cord. The currents that emerge from this calculation only depend on the empirically determined pair of networks, while the velocities and pressure gradients scale with the model parameters.

\section*{Materials and Methods}

\paragraph{Experimental microcosms}
\quad \newline
Cultures of \emph{P. velutina} (DC.) Parmasto were maintained in the Department of Plant Sciences, University of Oxford. The fungus was grown on 2\% malt agar (2\% no. 3 agar, 2\% malt extract, Oxoid, Cambridge, UK) at $22\pm 1 ^{\circ}\textrm{C}$ in darkness in a temperature-controlled incubator. To create wood inoculae, $1 \textrm{cm}^3$ autoclaved beech (\emph{Fagus sylvatica}) blocks (Bagley Wood sawmill, Kennington, UK) were placed on top of \emph{P. velutina} mycelium in the agar culture plates and incubated at $22^{\circ}\textrm{C}$, to allow penetration of the blocks by hyphae. Inoculated wood blocks were placed on a compressed bed of 33\% sterile white sand, 50\% sterile black sand, and 17\% water by weight in a 24-cm square culture dish. Two inoculated blocks were placed side-by-side in the center of each dish and allowed to grow at $21\pm 0.5 ^{\circ} \textrm{C}$ in the dark.

\paragraph{Producing digital networks from the experimental microcosms}
\quad \newline
The growing mycelium was photographed every three days, and the sequence of images was manually marked to record the location of nodes or junctions, as well as the presence or absence of edges. The cords were not sufficiently well resolved to make direct measurements of their diameter from the digitized images. However, the reflected intensity, averaged over a small user-defined kernel at either end of the cord, correlated well with microscope-based measurements of cord thickness. The observed relationship between image intensity and thickness was therefore used to estimate cord thickness across the mycelium (linear regression, $r2 = 0.77$, $df = 195$, $p < 0.0001$). This calibration was used to estimate the width of cords, while the volume of the mycelium was calculated by assuming that the cords were cylindrical (Bebber \emph{et al}. 2007; Ficker \emph{et al}. 2007, 2008; Jarrett \emph{et al}. 2006).

Three duplicate experiments, photographed over 36 days, were used to generate the results discussed in this paper. In each case the inoculum and resource units were represented as a single node, as the internal mycelial organization was not visible. Estimates for the diameter of cords range from 48 $\mu$m to 480 $\mu$m, although fine hyphae and cords smaller than 100 $\mu$m are likely to be missing from the digitised network.

\begin{figure}[htbp]
\begin{center}
\includegraphics[width=13cm]{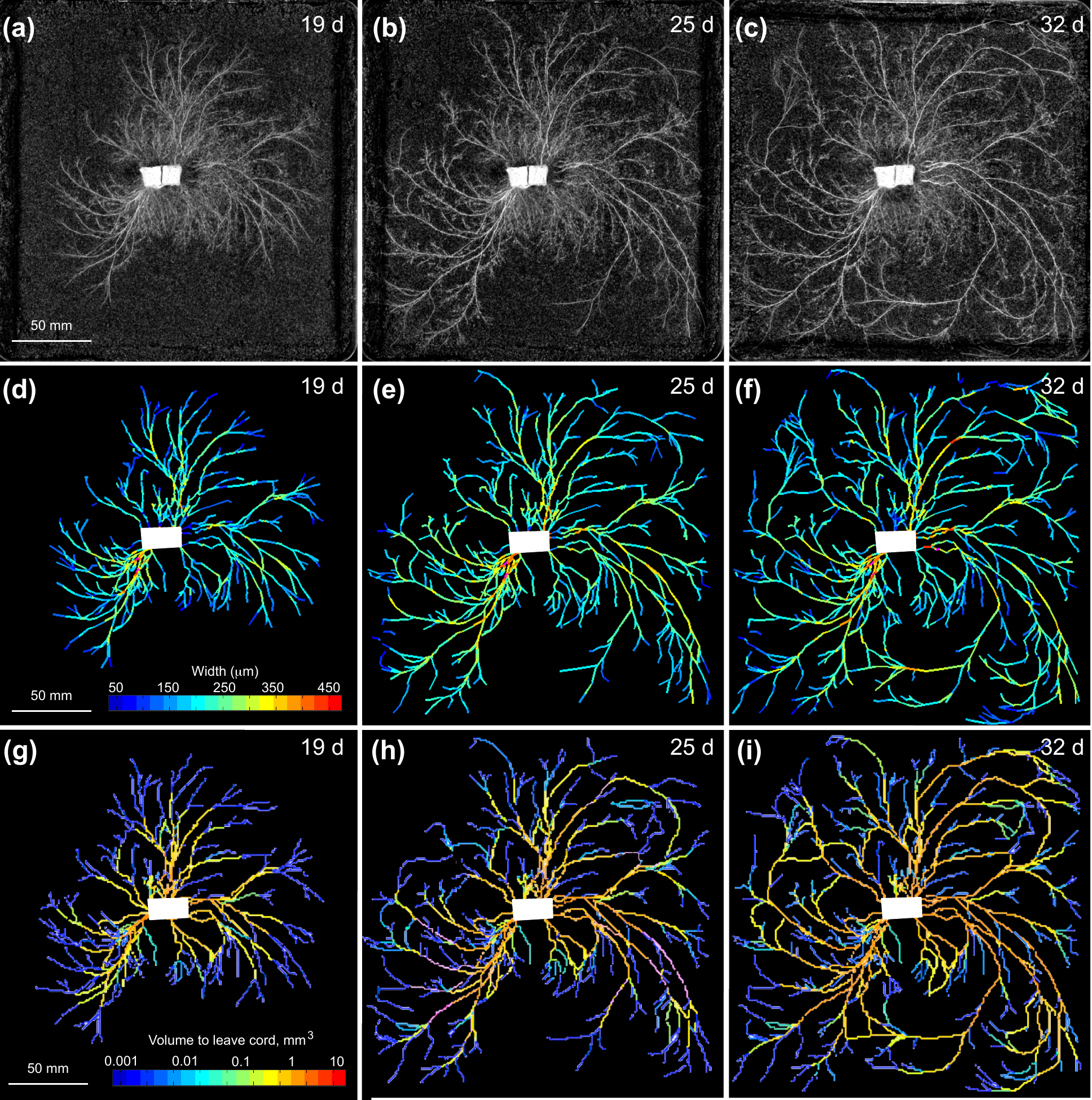}
\caption{\textbf{Network development and predicted currents in \emph{P. velutina}.} Images (a) to (c) show network development in \emph{P. velutina} after 19 days, 25 days and 32 days. The image intensity of cords was used to estimate their thickness, enabling the production of the weighted, digitized networks (d) to (f). These are colour-coded to show the estimated thicknesses of all sections of all edges. Images (g) to (i) are colour coded according to the total volume that has passed through each cord, as calculated by using the time-lapse model.}
\label{photos_to_predicted_currents}
\end{center}
\end{figure}

\section*{Results}
The total volume of the networks increased in an approximately linear fashion, so the proportional growth dropped significantly over time. The mean rate of increase over the first 21 days was $0.61 \textrm{mm}^{3}  \textrm{h}^{-1}$ in the first experiment, $0.51 \textrm{mm}^{3}  \textrm{h}^{-1}$ in the second experiment and $1.14 \textrm{mm}^{3}  \textrm{h}^{-1}$ in the third experiment. The total number of nodes and cords also increased in an approximately linear fashion, with around two cords and two nodes appearing every hour. The number of tips increased at about half that rate. In two of the experiments the total recorded volume of the network eventually decreased, and in all cases growth significantly slowed after 24 days. The time-lapse model was only applied while the fungi continued to grow at a rate of at least $0.1 \textrm{mm}^{3}$ per hour: a period of 21 days in experiment one, and 27 days in experiments two and three.

\paragraph{Correlation between the cross-sectional area of cords and topological traits}
\quad \newline
To assess the relationship between the topological organisation of the network and the cross-sectional area of the cords, we used the uniform model, with unit current at each tip and a constant conductance per unit length, to calculate a current for each cord. When using the uniform model, the network at each time point is effectively an independent experiment, and in all cases the calculated currents were correlated with the measured areas. Over the set of all networks, the mean value for the Spearman's rank correlation coefficient between current and cross-sectional area was $0.46 \pm 0.09$. When we considered the correlation between current and cross-sectional area for the complete set of edges (pooling the data from all the measured networks), the value of $\rho$ was $0.40$.

\begin{figure}[htbp]
\begin{center}
\includegraphics[width=13.5cm]{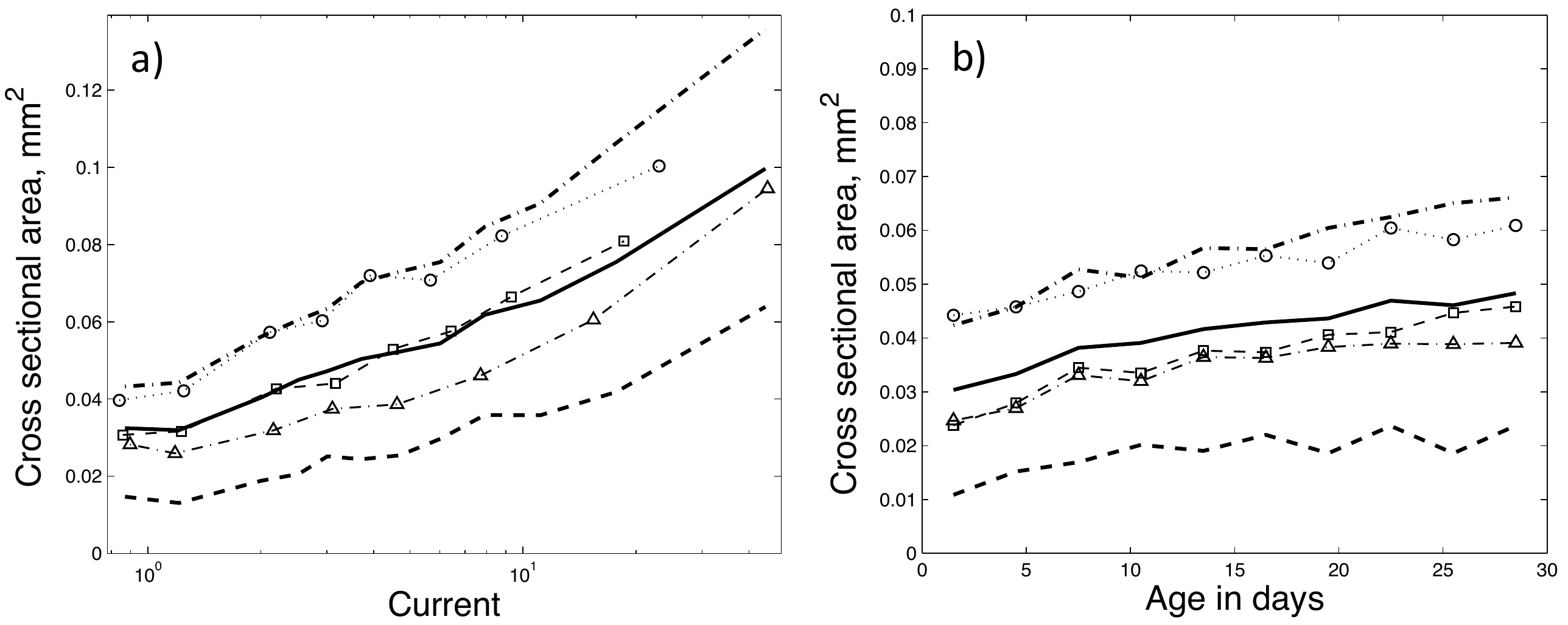}
\caption{\textbf{Correlation between cord cross-sectional area and (a) the current predicted by the uniform model, or (b) cord age.} Both graphs indicate the mean over all networks (thick solid line), and first and third quartiles over all networks (thick dashed lines). Other markers indicate the mean values for individual experiments. \newline \newline
\textbf{a)} There was a positive correlation between cross-sectional area and the current in a cord, where current is calculated by applying a unit current to each tip, and we assume constant conductance per unit length. The graph was produced by taking data from all time steps and partitioned it into bins. The first data point marks the mean cross-sectional area for cords with a current of one or less (as was the case for 30\% of cords), while the second cross marks the mean cross-sectional area for cords with a current between one and two (as was the case for 25\% of cords). The remaining cords were partitioned into bins of equal size according to the calculated current, and each cross marks the mean current and mean area of one of these bins. \newline \newline
\textbf{b)} Although most cords thicken over time, there was only a weak correlation between the age of a cord and its cross-sectional area. Note that the difference in values on the $y$-axis between young and old edges is small compared to the spread within each age group. Current is a better predictor of cross-sectional area.}
\label{age_and_top_current_v_area}
\end{center}
\end{figure}

Cords that were closer to the inoculum tended to be thicker, and older edges also tended to be thicker. However, current was a significantly better predictor of area than either distance or age (Fig. \ref{age_and_top_current_v_area}). More specifically, over the set of all networks the mean value of $\rho$ between the distance to the inoculum and the cross-sectional area of each cord was $-0.37 \pm 0.12$. When the data from all the networks was pooled, the value of $\rho$ was $-0.31$. Over the set of all networks (excluding those from the first time step, where, to our knowledge, all edges are the same age) the mean value of $\rho$ between the age and cross-sectional area of the cords was $0.41 \pm 0.14$. When the data from all the networks was pooled, the value of $\rho$ was $0.21$. As the relationship between current and area was reasonably consistent over time and over the three data sets (Fig. \ref{age_and_top_current_v_area}), it is possible to use this relationship to predict the size of cords given nothing more than the topology of a fungal network.

\begin{figure}[htbp]
\begin{center}
\includegraphics[width=8cm]{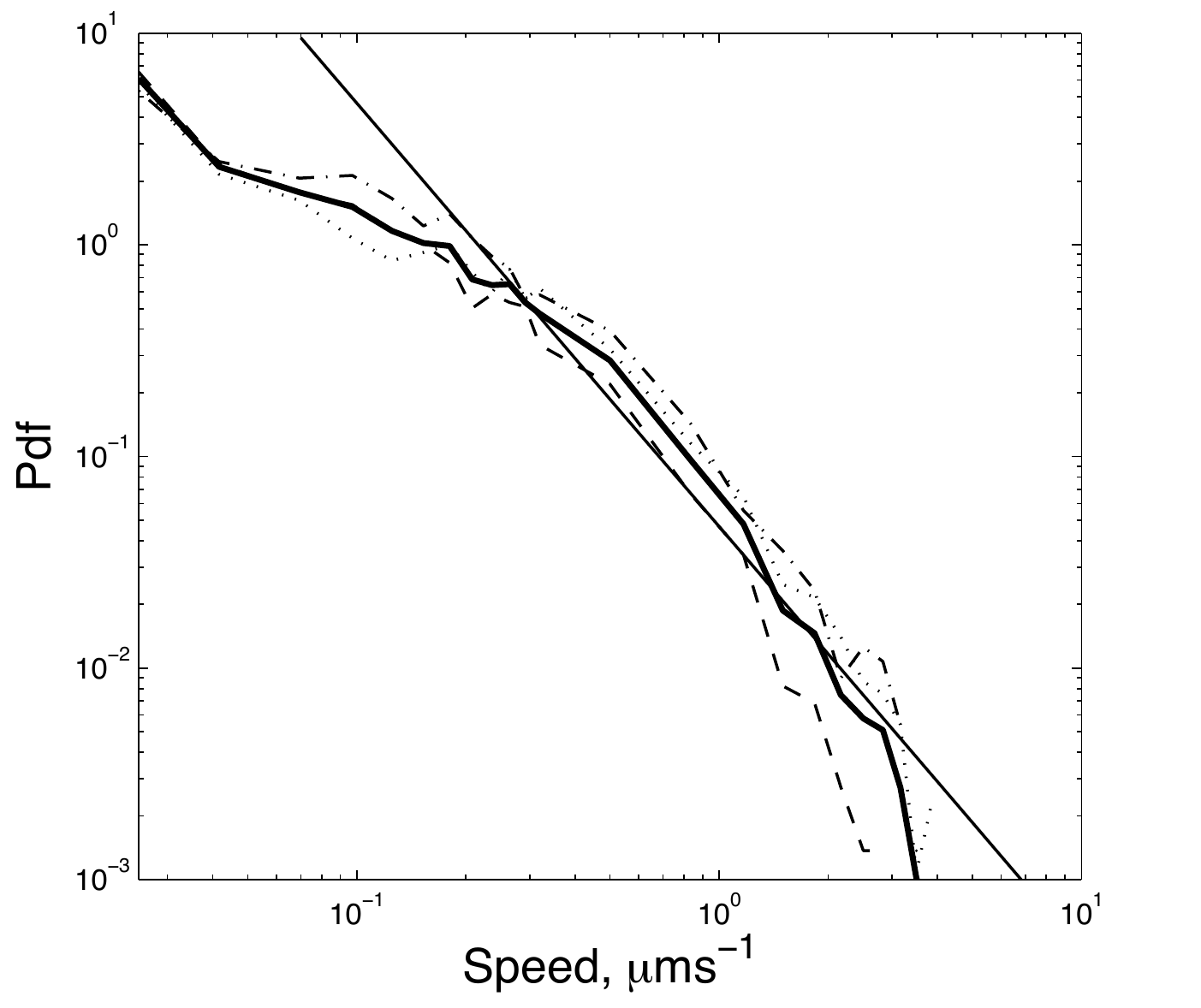}
\caption{\textbf{Log-log plot of the probability density function (pdf) of the predicted speed of flow, calculated using the time-lapse model.} The thick line indicates the pdf for the pooled data, and the dashed curves indicate the pdfs for individual experiments. The values for the speeds were calculated with the additional assumption that half the cross-sectional area of each cord was occupied by the interior of the vessels that carry mass flows. For comparison, the straight line represents the pdf for a branching tree of uniform thickness with $2^{n}$ `leaves' carrying mass flows of velocity $0.07 \mu \textrm{m s}^{-1}$, supplied by $2^{n-1}$ cords carrying mass flows of velocity $0.14 \mu \textrm{m s}^{-1}$, supplied by $2^{n-2}$ cords carrying mass flows of velocity $0.28 \mu \textrm{m s}^{-1}$, and so on down to a single trunk. The mean speed in such a branching tree would be massively dominated by the few, very fast cords. The pdf obtained from the time-lapse model decays more rapidly than a straight line, which indicates that the mean speed in the fungal network is less dominated by the exceptionally large speeds.}
\label{PDF_of_speeds}
\end{center}
\end{figure}

\paragraph{Distribution of currents and speeds}
\quad \newline
Both the uniform model and the time-lapse model indicate that many cords carried small currents while a few cords carried much larger currents. Furthermore, the cords that carry exceptionally large currents are sufficiently prevalent to dominate the mean current, so the majority of cords carrying a fraction of the mean current. The distribution of predicted speeds was similar, with many small velocities and a few much larger velocities (Fig. \ref{PDF_of_speeds}).  The speeds were calculated using the time-lapse model, with the additional assumption that half the cross-sectional area of each cord was occupied by the interior of the vessels that carry mass flows ($\lambda = 0.5$). Over all time steps and all experiments, 36\% of edges carry current at a speed greater than $0.1 \mu \textrm{m s}^{-1}$, 11\% of edges carry current at a speed greater than $0.5 \mu \textrm{m s}^{-1}$ and 4\% of cords carry current at a speed greater than $1 \mu \textrm{m s}^{-1}$. However, it should be noted that because the imaging process does not capture the growth of fine hyphae, these speeds represent a minimum estimate of the velocity of translocation.

\begin{figure}[htbp]
\begin{center}
\includegraphics[width=8.4cm]{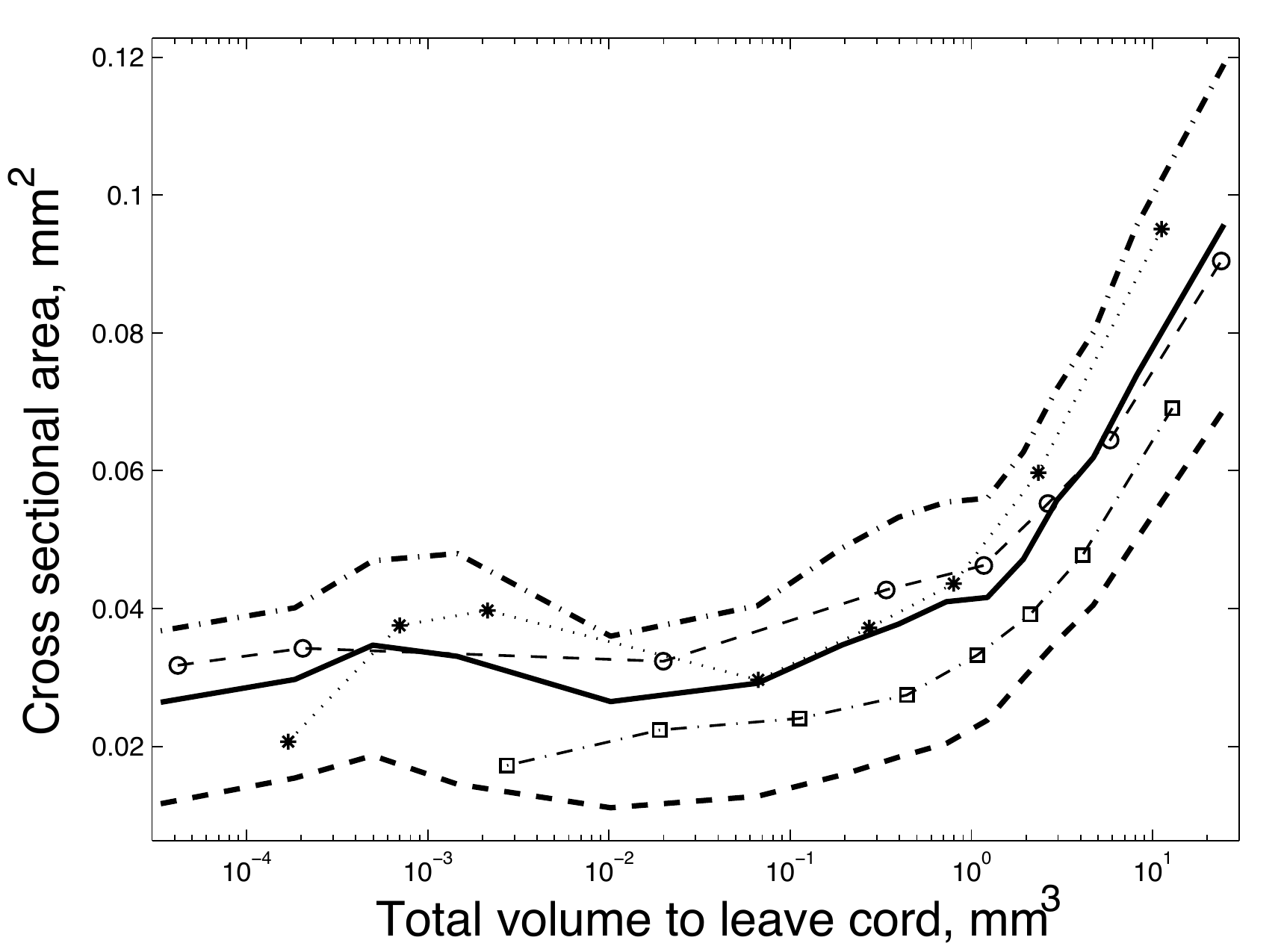}
\caption{\textbf{Correlation between cross-sectional area and the total volume to pass through each cord.} The data from all experiments and all time steps were partitioned into bins according to the total volume of fluid that had passed through each cord over its history up to each time point. The mean area (thick solid line) and first and third quartile values (thick dashed lines) have been plotted for each bin. The other markers indicate the means for individual experiments.}
\label{total_volume_v_area}
\end{center}
\end{figure}

 \paragraph{Correlations between area and the total volume passing through cords}
\quad \newline
At each time step the total volume to pass out of each cord was calculated over its history to date. There was a strong, positive correlation between the total volume flowing through a cord and its cross-sectional area (Fig. \ref{photos_to_predicted_currents} and \ref{total_volume_v_area}). The Spearman's rank correlation coefficient $\rho$ was 0.50, 0.56 or 0.57, and $\rho = 0.51$ for the pooled data. The vast majority of cords had a volume between $0.1$ and $1 \textrm{mm}^{3}$, and if we select a random edge at a random point in time, there was a $56\%$ chance that the total volume to leave the cord was greater than the volume of the cord itself.

\begin{figure}[htbp]
\begin{center}
\includegraphics[width=13.5cm]{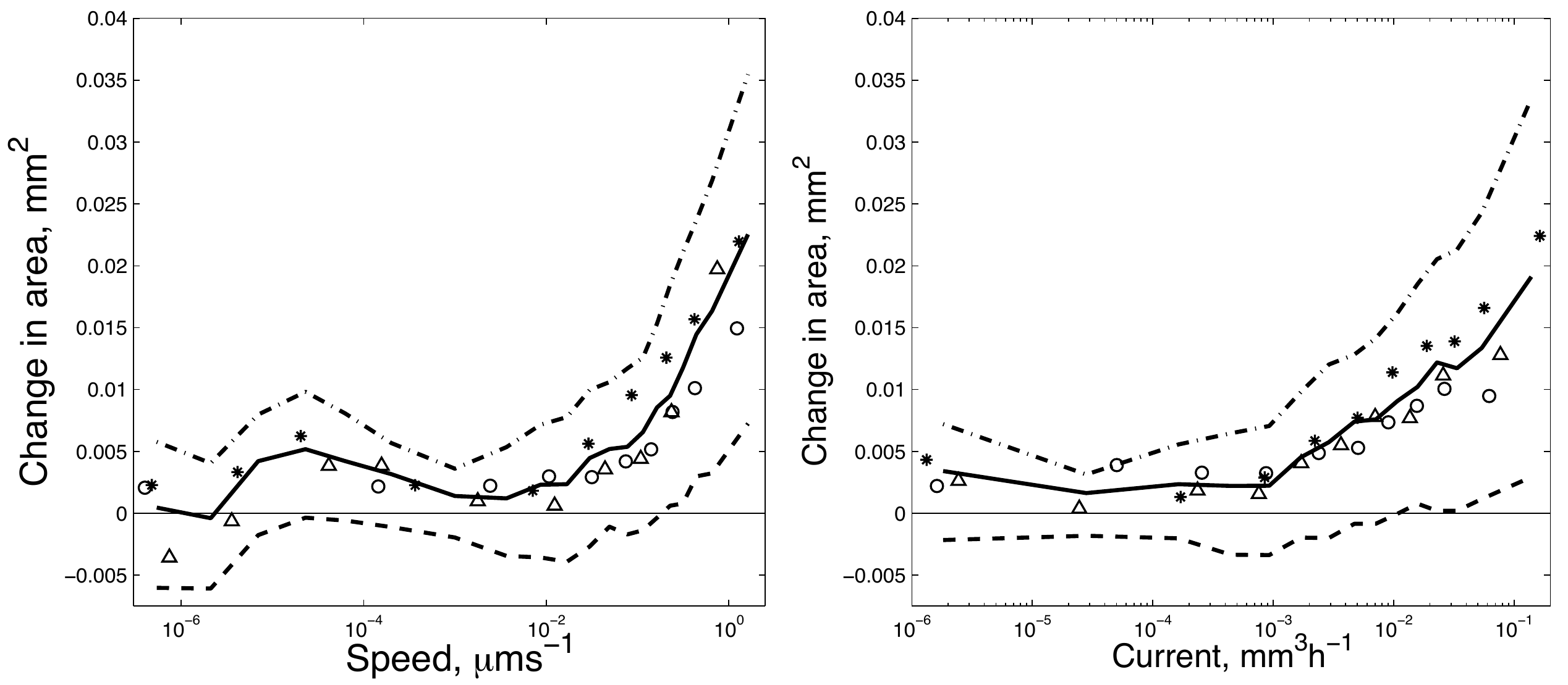}
\caption{\textbf{Correlation between the change in cross-sectional area and the predicted flow.} The graphs were produced by using the time-lapse model to calculate the current in each cord. The flux density or speed of flow was calculated by assuming that the interior of the vessels carrying the mass flows occupies half the total cross-sectional area. These graphs were produced by partitioning the data from all experiments and all time steps into ten bins of equal size according to the calculated speed or current. Each point on the curve indicates the mean speed (a) or mean current (b), plotted against the mean change in cross-sectional area for one of these bins. Thick solid lines indicate the mean over all experiments, and thick dashed lines indicate the first and third quartiles over all experiments. Other markers indicate the mean values for individual experiments. }
\label{speed_and_current_v_change}
\end{center}
\end{figure}

\paragraph{Correlations between speed of flow and changes in cross-sectional area}
\quad \newline
Cords that were predicted to carry a high velocity current were significantly more likely to increase in size than cords with a low velocity current (Fig. \ref{speed_and_current_v_change}a). Spearman's rank correlation coefficient $\rho$ between speed and change in area was 0.34, 0.28 or 0.34, and $\rho = 0.33$ for the pooled data. There was also a positive correlation between current and change in cross-sectional area (Fig. \ref{speed_and_current_v_change}b), with $\rho$ equal to 0.26, 0.20 or 0.32, with $\rho = 0.28$ for the pooled data.
Thicker cords tend to carry greater current, but this is to expected precisely because thicker cords have greater conductance. However, we also found that given a pair of
equally thick cords, the cord that is predicted to carry a greater current is the one that is more
likely to thicken (see SI Section 4, Fig. S2).

\section*{Discussion}

\paragraph{Growth, water uptake and mass-flows are coupled}
\quad \newline
If there is growth in a region, there must be a flow of material into that region (see Fig. \ref{growth_diagrams_1}b). When a fungus grows from a central inoculum across an inert substrate (e.g. Tlalka \emph{et al}., 2002; Tlalka \emph{et al}., 2003; Tlalka \emph{et al}., 2008), any volume for tip growth that is not derived from contracting regions must be acquired by transport from the inoculum.  In the microcosms under investigation here, \emph{P. velutina} was allowed to grow over sand that may have absorbed some moisture, so it is possible that water uptake occurred at locations other than the inoculum. However, cords have waxy coatings that insulate them from the environment (Cairney, 1992; Jennings, 1987; Rayner \emph{et al}., 1991). Our account of growth-induced mass flow indicates an advantage of insulated cords. By reducing the local uptake of water, insulation increases the distance between the sites of water uptake (source) and the regions of growth (sink), thereby increasing the scale of growth-induced mass flows (Banavar \emph{et al}., 1999; Banavar \emph{et al}., 2002; Dreyer \& Puzio, 2001).

\paragraph{Transport velocities are determined by the network architecture}
\quad \newline
The network architecture will affect the velocity of growth-induced mass flows. At one extreme, consider a binary branching tree, where at every step, every tip branches to produce two offspring, and there is no anastomosis or cord-thickening. At any given moment each generation of cords must carry the same total current out towards the tips, where the speed of flow is equal to the rate of tip growth. It follows that if mothers and daughters have the same cross-sectional area (as is the case for individual hyphae), the velocity of flow in a vessel with $n$ generations of offspring will be $2^{n}$ times greater than the rate of tip growth. In principle, this explains how water uptake and growth can induce mass flows at speeds that are orders of magnitude greater than the rate of individual tip growth, and the resulting distribution of velocities is illustrated in Fig. \ref{PDF_of_speeds}.

At the other extreme, a constant flow rate is predicted in transport systems like the xylem in plants, where vascular bundles form a branching hierarchy, but the total cross-sectional area is preserved at every stage  (Savage \emph{et al}., 2008; Sherman, 1981). Our analysis suggests that the distribution of velocities across the mycelium is closer to the former case (Fig. \ref{PDF_of_speeds}), with many cords carrying low velocity mass flows, while some cords carry high velocity mass flows. However, we cannot yet provide a full quantitative analysis of the predicted velocities as our imaging techniques do not have sufficient resolution to map the fine hyphae that fan out ahead of the developing cords. Thus the actual volume increase at one of the notional `tips' characterised here may well be several-fold greater than the volume of the terminal cord that we can measure.

With this caveat, the distribution of predicted currents should reflect the actual currents to the extent that the regions with large amounts of sub-resolution hyphal growth are also regions where the cords develop and thicken. For example, if we assume that $0.1 \textrm{mm}^{3}$ of fine hyphae grows out of each tip each day, the predicted growth-induced currents would approximately double. This estimate of the volume of fine hyphal growth corresponds to each tip growing a fan of $10 \mu \textrm{m}$ thick hyphae, which cover an area of $10 \textrm{mm}^2$ per day. If this coarse estimate of the quantity of fine hyphal growth is accurate, the volume of the digitised cords is only 50-60\% of the total fungal volume.

To provide a more accurate prediction of the maximum velocities
that would be produced if the fluids within fungi simply responded to growth by following the path of least resistance,
we would need better morphological information on the number and size of the conducting vessels at each stage of network development. Using the simplifying assumptions that the diameter of the vessel hyphae is constant ($12 \mu \textrm{m}$), but the number of conduits scales with the area of the cord and that 50\% of the cross-sectional area of the cord comprises conducting vessels, the time-lapse model predicts that 4\% of cords carried velocities greater than $1 \mu \textrm{m s}^{-1}$.
This 4\% of cords that carry the most current approximately corresponds to the major, or most visible, cords in the network.

It may be more realistic to suppose that  growth-induced mass flows are carried through vessels that only occupy 10\% of the cords' cross-sectional area (rather than the previous estimate of 50\%). If we make this assumption, and suppose that  $0.1 \textrm{mm}^{3}$ of fine hyphae grows out of each tip each day, our time-lapse model predicts that 4\% of cords would carry mass flows with velocity greater than $10 \mu \textrm{m s}^{-1}$. This is comparable to the kinds of velocities observed experimentally in the major cords of \emph{P. velutina} and greater than those reported for \emph{Schizophyllum commune} (Olsson \& Gray 1998; Tlalka \emph{et al}. 2002; Brownlee \& Jennings 1982; Connolly \& Jellison 1997; Lindahl \emph{et al}. 2001; Thompson \emph{et al}. 1985, 1987). However, as the calculations are critically dependent on the network architecture, estimation of the precise contribution of growth to mass-flows would require both detailed volume measurements and very precise velocity measurements on individual microcosms, that is currently beyond our technical capability.
Direct comparisons between predicted and measured velocities may be more feasible in the smaller and non-corded networks formed by species such as \emph{Neurospora crassa} (Lew 2005).

\paragraph{Cords with high current or velocity increase in area}
\quad \newline
Mass flows are required to maintain a sufficient supply of nutrients throughout the mycelium, but there are limits to the current that can pass along an individual hypha. High velocities require high pressure gradients, and increasing the velocity of flow means that greater amounts of work must be done by the fungi to overcome viscous drag. Thickening cords and the formation of high conductivity, aseptate channels may represent effective responses to these challenges.

We have observed a characteristic relationship between the total current that has passed through a cord, and the thickness of that cord (Fig. \ref{photos_to_predicted_currents} and \ref{total_volume_v_area}). We have also observed a correlation between the currents and the flux densities predicted by our model and the extent to which cords thicken over time (Fig. \ref{speed_and_current_v_change}). Given the further assumption that \emph{P. velutina} has adapted to reduce the work needed to overcome viscous drag, we should expect to see preferential thickening of the high current cords. This is because, where there is a distribution of currents, significantly greater energy savings can be made by thickening the high current cords (as opposed to thickening the low current cords).

There is an element of positive feedback inherent in these observations, as any differential thickening of two parallel transport pathways will automatically increase flow through the cord with greater hydraulic conductivity. However, while we should expect that larger cords will carry greater currents (precisely because they have greater conductance), we have also found that given a pair of equally thick cords, the cord that is predicted to carry a greater current is the one that is more likely to thicken (see SI Section 4, Fig. S2).

\paragraph{Conclusion}
\quad \newline
In conclusion, we note that the incompressibility of the fluids within fungi ensures that there is a rapid global response to local fluid movements. Furthermore, velocity of fluid flow is a local signal that can convey quasi-global information about the role of a cord within the mycelium. We have found a correlation between the thickening of cords and the speeds or flux densities predicted by our model (Fig. \ref{speed_and_current_v_change}a and S2a). Similarly, there was a positive correlation between predicted current and the thickening of cords (Fig. \ref{speed_and_current_v_change}b and S2b). This is consistent with the plausible assumption that \emph{P. velutina} has evolved to reduce the work needed to overcome viscous drag, as significantly greater energy savings can be made by preferentially thickening the high current cords.

The speeds predicted by our model are consistent with experimental data, and the pressure gradients required to produce the predicted flows are very modest (see SI Section 3). Furthermore, contrary to previous analyses, we suggest that intrahyphal concentration gradients are not strictly necessary for the production of mass flows. The uptake of water and the maintenance of turgor pressure require an osmotic gradient between the hyphae and their environment, but the incompressibility of aqueous fluids ensures that there will be a mass flow from the sites of water uptake to the sites of growth, regardless of the concentration gradients within the mycelium itself. We also suggest that local responses to flux density and nutrient concentration might govern the development of these remarkable self-organizing, efficient, adaptive, growing transport networks.

\paragraph{Acknowledgements}
\quad \newline
LLMH thanks the EPSRC for financial support. NSJ thanks the EPSRC and BBSRC. EL thanks the EPSRC.
PKM was partially supported by a Royal Society Wolfson Merit Award. MDF thanks the BBSRC and NERC
for financial support, and J. Lee for technical assistance.

\newpage

\section*{Growth-induced mass flows in fungal networks: \\Supplementary Information}

\vspace{1cm}
\section{Comparison of models}

\begin{center}
\begin{figure}[htbp]
\includegraphics[width=13.2cm]{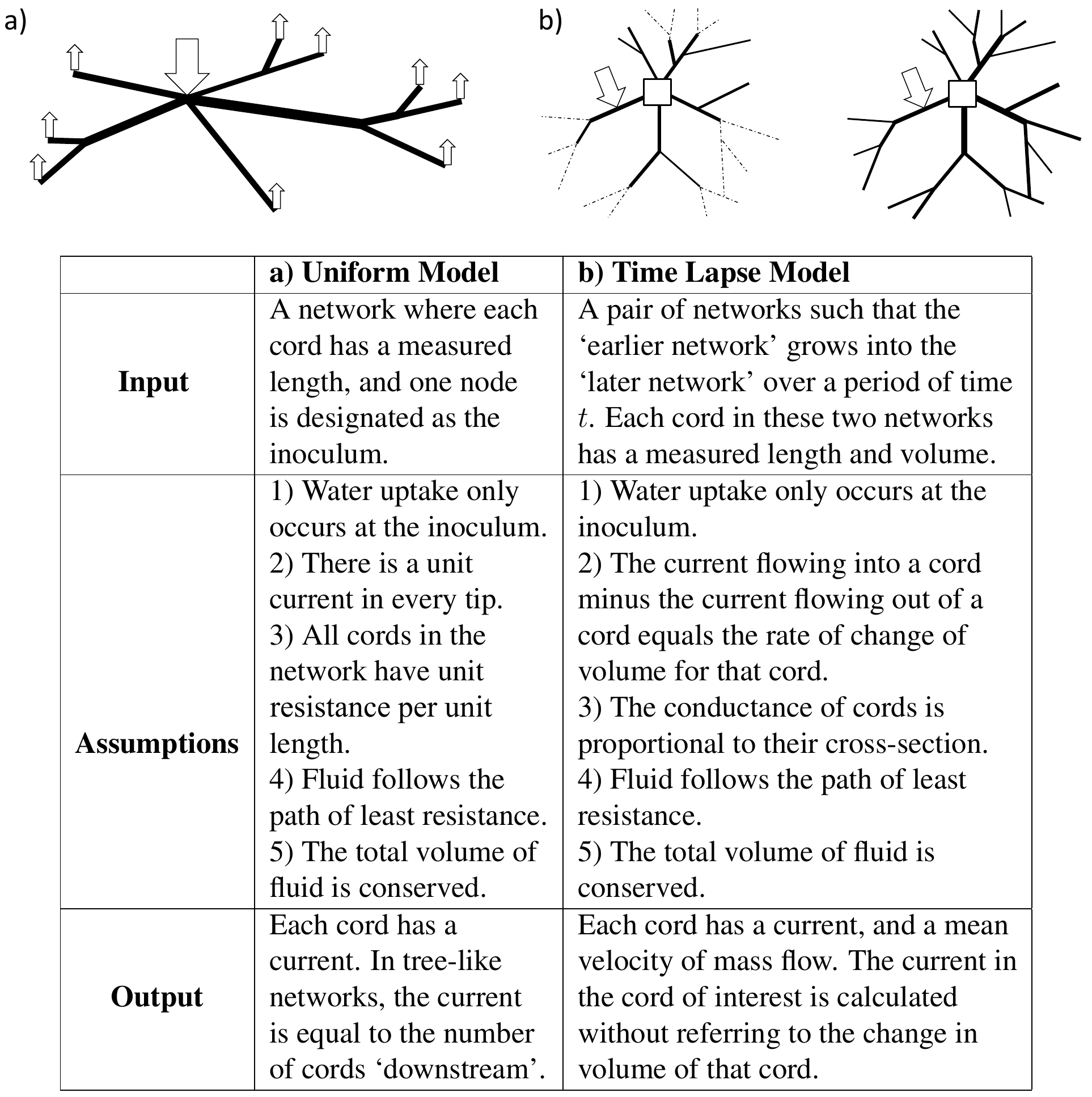}
\caption{\textbf{Models of growth-induced mass flow.}}
\label{models}
\end{figure}
\end{center}
\clearpage

\section{Parameters and solutions of the time lapse model}

\paragraph{Calculating the pressure at each node}
\quad \newline
Recall that the current in a cord is equal to the pressure drop times the conductance of the cord. It follows that at each node $i$
\begin{equation}
\sum_{j} ( p_{i} - p_{j} ) C_{ij} = q_{i},
\label{p_C_and_q}
\end{equation}
where $p_{i}$ is the pressure at node $i$, $q_{i}$ is the net current flowing out of node $i$ and $C_{ij}$ is the conductance of the cord between nodes $i$ and $j$. We can summarise the relationship expressed by Equation (\ref{p_C_and_q}) by defining the following symmetric matrix \textbf{A}:

\begin{eqnarray}
A_{ii} & = & \sum_{k} C_{ik} \quad \textrm{ and} \nonumber \\
A_{ij} & = & -C_{ij}. 
\label{A_def}
\end{eqnarray}

Letting $\overline{p}$ be the vector form of the pressures $p_{i}$ and $\overline{q}$ the vector form of the currents $q_{i}$, we know by Equation (\ref{p_C_and_q}) that $\overline{p}$ satisfies the equation 

\begin{equation}
\textbf{A} \overline{p} = \overline{q}.
\label{linear_system}
\end{equation}

Since each row of \textbf{A} sums to zero, \textbf{A} has no inverse and we cannot uniquely determine 
$\overline{p}$. However, it is the pressure differences which are of interest, not the absolute value of $\overline{p}$. We are therefore free to fix the pressure at any one node. Once we have done this, Equation (\ref{linear_system}) uniquely determines the pressure at every other node. More specifically, suppose that our network contains $N$ nodes. In that case Equation (\ref{linear_system}) represents a system of $N$ linear equations in $N$ unknowns, but as each row and column sums to zero, the $N$'th equation is a linear combination of the other $N-1$ equations. Setting $p_{N} = 0$ gives us a system of linear constraints on the values for $p_{1}, \ldots, p_{N-1}$, namely

\begin{equation*}
\left(\begin{array}{cccc}
A_{11} & \cdots & A_{1(N-1)} & A_{1N} \\
\vdots & \ddots & \vdots & \vdots \\
A_{(N-1)1} & \cdots & A_{(N-1)(N-1)} & \\
A_{N1} & \cdots & & A_{NN} 
\end{array}\right)
\left(\begin{array}{c}
p_{1} \\
\vdots \\
p_{N-1} \\
0
\end{array}\right) = 
\left(\begin{array}{c}
q_{1} \\
\vdots \\
q_{N-1} \\
q_{N}
\end{array}\right).
\end{equation*}

This is equivalent to the following  system of linear constraints on the values for $p_{1}, \ldots, p_{N-1}$,

\begin{equation}
\left(\begin{array}{ccc}
A_{11} & \cdots & A_{1(N-1)}  \\
\vdots & \ddots & \vdots \\
A_{(N-1)1} & \cdots & A_{(N-1)(N-1)}
\end{array}\right)
\left(\begin{array}{c}
p_{1} \\
\vdots \\
p_{N-1}
\end{array}\right) = 
\left(\begin{array}{c}
q_{1} \\
\vdots \\
q_{N-1}
\end{array}\right)
\label{reduced_linear}
\end{equation}
plus an additional equation 
\begin{equation}
A_{(N-1)1}p_{1} + A_{(N-1)2}p_{2} + \ldots + A_{(N-1)(N-1)}p_{N-1} = q_{N}.
\label{redundant_equation}
\end{equation} 

Because the rate of water uptake is equal to the rate of growth, the total in-current is equal to the total out-current. In other words, $\sum q_{i} = 0$. We therefore know that $q_{N}=-q_{1} - \ldots - q_{N-1}$. We can use Equation (\ref{reduced_linear}) to eliminate each of the $q_{i}$ (as $q_{i} = A_{i1}p_{1} + A_{i2}p_{2} + \ldots + A_{i(N-1)}p_{N-1}$), and this shows that Equation (\ref{redundant_equation}) is redundant. In other words, given that the total in-current is equal to the total out-current, Equation (\ref{reduced_linear}) is sufficient to determine our solution. It contains $N-1$ independent linear equations in $N-1$ unknowns, which uniquely determines the
values $p_{1}, \ldots, p_{N-1}$. 

\paragraph{The parameters of the time lapse model}
\quad \newline
 We require the parameter $\delta$ because newly forming cords must have some conductance, or else they would not be connected to the rest of the network, and their growth could not induce currents. The precise value of $\delta$ is not significant, but the calculation of currents requires that all cords have a non-zero conductance. Alternatively, we could remove the parameter $\delta$ by supposing that the conductance of cord $ij$ is $\frac{\sigma(u_{ij} + v_{ij})}{2l_{ij}}$. Since an extension of a cord can be represented as a new cord, the length of cords does not change. Hence this formula is  equivalent to supposing that the cross-sectional area of cord $ij$ is half way between the cord's cross-section in the initial network and its cross-section in the resulting network. However, in that case cords that increase in size would carry more current than cords that do not, precisely because in this alternate model, cords that become large are assigned a larger conductance. This is undesirable (and not the method we employ), because we want an unbiased estimate of current to correlate with changes in area. We could also have avoided the parameter $\delta$ by saying that newly forming cords are not part of the network, but in that case we would have to devise an additional algorithm for assigning out-current to the cords from which the new cords grow: something we `get for free' by using our model with $\delta$.

The parameter $\sigma$ specifies the conductance per unit area for each cord in the network. The value of $\sigma$ does not affect the calculated currents, as it is the relative conductance of cords that determines the distribution of currents. However, the pressure gradients predicted by this model will be inversely proportional to $\sigma$.

Varying the parameter values $\sigma$ and $\delta$ does not significantly affect our calculation of the currents induced by growth. However, we do need to make a significant assumption concerning the relationship between conductance and cross-sectional area. We assume that conductance is proportional to cross-sectional area, but this assumption is only of consequence when the network provides a number of alternate routes between the site of water uptake and the site of growth. Because our boundary conditions are fixed in terms of currents, the current in part of a branching tree will not depend on the conductance of the network. For example, if there is a certain current flowing out of the tips of a branching tree, the specified current has to flow through a given sequence of cords regardless of their conductance, as in the absence of loops is only only one route from trunk to tip. In practice, the fungal networks we are studying are composed of branching trees connected to a net-like core. Within the net-like structure more current will tend to flow through the larger cords, as fluids naturally follow the `path of least resistance'.

\vspace{2cm}

\section{Pressure gradients and wall shear stress}

\paragraph{Viscosity, velocity and laminar flow}
\quad \newline
To assess whether the flows in fungi are laminar, we first estimate the Reynolds number (Re). This is defined as the ratio of inertial to viscous forces (Nobel 1991), that is
\begin{equation}
\mathrm{Re} = \frac{\rho v d}{\eta},
\end{equation}
where $\rho$ is the density of the fluid (approximately equal to that of water: $1 \mathrm{g ml}^{-1}$), $v$ is the mean velocity of the fluid, $d$ is the diameter of the hypha or transport vessel, and $\eta$ is the dynamic viscosity  of the fluid.  The viscosity of cytoplasm is reported to be similar to that of water, $1  \mathrm{g s^{-1} m}^{-1}$ (see supplementary reference Fushimi \& Verkman, 1991), but the fluids within fungi could plausibly be as viscous as 1.5 M sucrose solution, which has a viscosity of $7  \mathrm{g s^{-1} m}^{-1}$  (Bancal \& Soltani 2002). Here we use a value of $\eta = 2 \mathrm{g s^{-1} m}^{-1}$.

The velocity of fluid flow and the diameter of the tubular vessels within the cords may vary considerably throughout the fungi, but even the most extreme plausible values yield a Reynolds number several orders of magnitude smaller than one (Lew 2005). This tells us that smooth, laminar flow is occurring (Nobel 1991; Lew 2005).

\paragraph{Pressure gradients and speed of flow}
\quad \newline
Suppose that a cord has a cross-sectional area $a$ and carries a current $f$. The mean speed of flow within the cord as a whole will be $f/a$. However, only a fraction $\lambda$ of the cross-sectional area of each cord will be occupied by the interior of the vessels that carry mass flows. Thus the mean speed within the vessels will be $f/\lambda a$. If we want to use our model to obtain estimates for the speeds of mass flow we need to choose an appropriate value for $\lambda$. Similarly, if we want estimates of the pressure gradients we need to choose a sensible value for the parameter $\sigma$. Here we assume that $\lambda = 0.5$, and that the tubes carrying mass flows all have an internal radius $r=6 \mu\textrm{m}$. In this case
\begin{equation}
v = \frac{f}{\lambda a},
\end{equation}
where $v$ is the mean velocity of mass flow, $f$ is the current through the cord and $\lambda a$ is the cross-sectional area through which the current passes. The Hagen-Poiseuille equation tells us that the pressure gradient $\frac{\textrm{d}P}{\textrm{d}x}$ must satisfy the equation
\begin{equation}
\frac{\textrm{d}P}{\textrm{d}x} \equiv \frac{\Delta P}{l} = f \frac{8 \eta}{n \pi r^{4}} = f \frac{\pi r^{2}}{\lambda a} \frac{8 \eta}{\pi r^{4}} = v  \frac{8 \eta}{r^{2}},
\label{pressure_speed_viscosity}
\end{equation}
where $\eta$ is the dynamic viscosity of the fluid,  $n$ is the number of tubes within the cord and $r$ is the radius of each tube.

\paragraph{Deriving pressure gradients from the time lapse model}
\quad \newline
To estimate the pressure gradients needed to drive the flows predicted by the time lapse model, we need estimates for the conductances of cords. Here we assume that half the cross-sectional area of each cord was occupied by transport vessels of internal radius of $6 \mu\textrm{m}$, and we also assume that the viscosity of the moving fluids was $2 \textrm{gs}^{-1} \textrm{m}^{-1}$ (Eamus 1985; Howard 1981; Lew 2005). By Equation (\ref{pressure_speed_viscosity}), these values give us the relationship 
\begin{equation*}
\frac{\textrm{d}P}{\textrm{d}x} \approx 4v \times 10^{-5},
\label{pressure_to_speed}
\end{equation*}
where $v$ is measured in $\mu\textrm{m s}^{-1}$, and $\frac{\textrm{d}P}{\textrm{d}x}$ is measured in $\textrm{bar cm}^{-1}$. Our estimate of conductance per unit area tells us that maintaining a velocity of $1 \mu\textrm{m s}^{-1}$  only requires a pressure gradient of around $4 \times10^{-5} \textrm{ bar cm}^{-1}$. This is  very small compared to the hydrostatic pressure of hyphae, which is about $4-5 \textrm{ bar}$ (Amir \emph{et al}. 1995b; Lew \emph{et al}. 2004; Lew 2005; Money 1997). Pressure gradients of this scale could plausibly be maintained over tens or even hundreds of meters. 

In \emph{Neurospora crassa} the cytoplasm moves forward with the growing tips at a rate of $0.2 - 0.5 \mu\textrm{m s}^{-1}$. Mass flows in the hyphae behind the tips typically reach $5 \mu\textrm{m s}^{-1}$, and currents as fast as $60 \mu\textrm{m s}^{-1}$ have also been directly observed (Lew 2005). In \emph{P. velutina} the highest reported velocities are around $900 \mu\textrm{m s}^{-1}$ (Wells \emph{et al}. 1995b), though obtaining accurate estimates of velocity is a major challenge.  Our estimate for the conductance of cords implies that maintaining a velocity as large as $900 \mu\textrm{m s}^{-1}$ requires a pressure gradient of around $0.04 \textrm{ bar cm}^{-1}$. This is a significant pressure gradient compared to the hydrostatic pressure of hyphae, and pressure gradients of this scale could only be sustained over a few tens of centimeters.

\paragraph{Deriving wall shear stresses from the time lapse model}
\quad \newline
Fluid flows induce wall shear stresses on the vessels within cords. A good estimate for the wall shear stress  $\tau$ can be obtained using the formula
\begin{equation}
\tau = \frac{4 \eta v}{r},
\label{stress_velocity}
\end{equation}
where $\eta$ is the dynamic viscosity of the fluid, $v$ is the mean velocity of fluid flow and $r$ is the radius of the vessels within cords (Sherman 1981). Using the previously indicated values for $\eta$ and $r$ tells us that
\begin{equation*}
\tau \approx v \times 10^{-3},
\end{equation*}
where the mean velocity $v$ is measured in $\mu \textrm{m s}^{-1}$ and the wall shear stress $\tau$ is measured in pascals or $\textrm{Nm}^{-2}$. By way of comparison, the wall shear stresses in mammalian arterial systems are in the range $0.2-2 \textrm{Nm}^{-2}$ (Kamiya \emph{et al}. 1984; Rodbard 1975). 

It is widely accepted that a local adaptive response to wall shear stress is a key mechanism that enables the optimisation of mammalian vascular systems (Kamiya \emph{et al}. 1984; Rodbard 1975; Sherman 1981). By analogy it is certainly plausible that hyphae could detect and respond to velocities of the order $100 - 1000 \mu\textrm{m s}^{-1}$, as we estimate that such currents would induce wall shear stresses of the order $0.1 - 1 \textrm{Nm}^{-2}$. It is less likely that fungi can detect the difference between much slower moving currents, as the corresponding changes in wall shear stress would be very small.

\newpage
\section{Further Results}

\begin{figure}[htbp]
\begin{center}
\includegraphics[width=13.5cm]{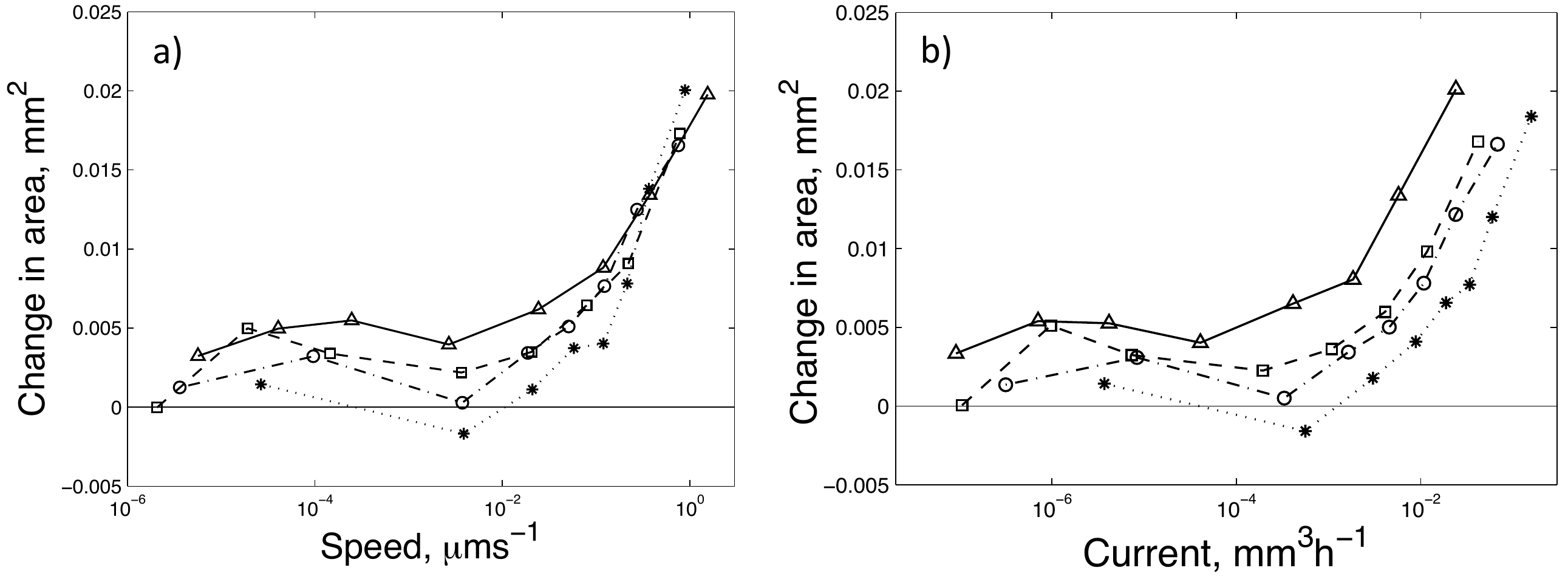}
\caption{\textbf{Correlation between the change in cross-sectional area and the predicted flow for edges of similar thickness.} The data from all experiments and all time steps were partitioned into four bins according to the thickness of the cords. The lines were generated by data from cords with cross-sectional area less than $0.02 \textrm{mm}^{2}$ ($\triangle$), cross-sectional area between $0.02 \textrm{mm}^{2}$ and $0.04 \textrm{mm}^{2}$
 ($\Box$), cross-sectional area between $0.04 \textrm{mm}^{2}$ and $0.06 \textrm{mm}^{2}$ ($\circ$) and cross-sectional area greater than $0.06 \textrm{mm}^{2}$ ($\ast$). Each of these bins was then subdivided into ten subsets of equal size, according to the calculated speed (a) or current (b). Each marker indicates the mean speed and mean change in cross-sectional area for one of these subsets. \newline \newline
 \textbf{a)} Regardless of cross-sectional area, there was a similar relationship between the speed of flow in a cord and the change in cross-sectional area. Note that fewer large cords have a very low speed (less than $10^{-2}$, say).\newline \newline
\textbf{b)} Larger cords tended to carry more current, but regardless of cross-sectional area, there was a similar correlation between predicted current and the measured change in area.}
\label{speed_and_current_v_change_by_area}
\end{center}
\end{figure}

\newpage
\section{Growth, Mass Flows and Nutrient Translocation}
\quad \newline
The currents calculated by the time-lapse model represent a minimal total flux, found by calculating the unique set of currents that account for the observed changes in cord volume, while minimising the work required to overcome viscous drag. A different distribution of currents could be established by additional transport mechanisms, though the conservation of volume is a constraint on all possible patterns of fluid flow. In particular, osmotic gradients between adjacent fungal vessels could produce flows that differ from the calculated minimum. Indeed, if transport were solely driven by apical mass flow, it would be difficult to account for simultaneous bi-directional movement and concurrent basal transport (Fricker \emph{et al}. 2007; Lindahl \emph{et al}. 2001; Olsson \& Gray 1998; Tlalka \emph{et al}. 2003,  2008). However, any currents beyond those predicted by our model necessarily require the fungi to do additional work.

Our central claim is that nutrients loaded into the mass-flow transport pathway will move towards the hyphal tips at a velocity that is partly determined by the volume of `downstream' growth, and the network architecture. However, to actually reach the tips, the nutrients need to move at a greater speed than the column of water that advances in tandem with the tips (see Fig. 1a). This will be achieved in part automatically at a slow rate by diffusion, but could be increased substantially by local evaporation at the tips, the movement of vesicles by motor proteins, or by other means of active transport. 

Finally, we note that as growth, mass-flow and nutrient transport are coupled, there may be an interesting interaction between nutrient availability,  control of branching and nutrient transport. It is well known that the rate of hyphal branching increases when tips encounter resource rich environments (Gow \& Gadd 1995). Turgor pressure and the build up of vesicles have both been implicated, but whatever the mechanism behind this response, differential branching rates may constitute a unique kind of foraging strategy. In resource-poor regions tips rarely branch, and consequently tip growth in resource-poor environments induces relatively low flux densities in the trailing hyphae. In resource-rich environments, where branching rates are high, flux densities will also be high. We speculate that regions of the mycelium that do not receive a sufficient supply of resources will regress, and that the resulting networks constitute an efficient response to the given resource environment.


\begin{thebibliography}{label}

\addcontentsline{toc}{section}{\refname}

\bibitem{amir2}
\textbf{Amir R, Levanon D, Hadar Y, Chet I. 1995a.} Factors Affecting Traslocation and Sclerotial Formation in \emph{Morchella esculenta}. \emph{Exp Mycol}. \textbf{19}: 61-70.

\bibitem{amir}
\textbf{Amir R, Steudle E, Levanon D, Hadar Y, Chet I. 1995b.} Turgor changes in \emph{Morchella esculenta} during translocation and sclerotial formation. \emph{Exp Mycol}. \textbf{19}: 129-136.


\bibitem{banavar}
\textbf{Banavar JR, Maritan A, Rinaldo A. 1999.} Size and form in efficient transportation networks. \emph{Nature}. \textbf{399}: 130-132.

\bibitem{banavar2}
\textbf{Banavar JR, Maritan A, Rinaldo A. 2002.} Supply-demand balance and metabolic scaling. \emph{Proc Natl Acad Sci USA}. \textbf{99}: 10506-10509.

\bibitem{bancal}
\textbf{Bancal P, Soltani F. 2002.} Source-sink partitioning. Do we need M\"{u}nch? \emph{J Exp Bot}. \textbf{53}: 1919-1928.

\bibitem{bebber}
\textbf{Bebber D, Hynes J, Darrah PR, Boddy L, Fricker MD. 2007.} Biological solutions to transport network design. \emph{Proc Roy Soc B}. \textbf{274}: 2307-2315.


\bibitem{boddy}
\textbf{Boddy, L. 1999.} Saprotrophic cord-forming fungi: meeting the challenge of heterogeneous environments.  \emph{Mycologia}. \textbf{91}: 13-32.

\bibitem{boswell}
\textbf{Boswell G, Jacobs H, Davidson FA, Gadd GM, Ritz K. 2002.} Functional consequences of nutrient translocation in mycelial fungi.  \emph{J Theor Biol}. \textbf{217}: 459-477.

\bibitem{boswell2}
\textbf{Boswell G, Jacobs H, Davidson FA, Gadd GM, Ritz K. 2003a.} A mathematical approach to studying fungal mycelia.  \emph{Mycologist}. \textbf{17}: 165-171.

\bibitem{boswell3}
\textbf{Boswell G, Jacobs H, Davidson FA, Gadd GM, Ritz K. 2003b.} Growth and function of fungal mycelia in heterogeneous environments.  \emph{Bull Math Biol}. \textbf{65}: 447-477.

\bibitem{boswell5}
\textbf{Boswell G, Davidson FA, Ritz K, Gadd GM, Jacobs H. 2007.} The development of fungal networks in complex environments.  \emph{Bull Math Biol}. \textbf{69}: 605-634.

\bibitem{brownlee}
\textbf{Brownlee C, Jennings DH. 1982.} Long distance translocation in \emph{Serpula lacrimans}: velocity estimates
and the continuous monitoring of induced perturbations. \emph{Trans Br Mycol Soc}. \textbf{79}: 43-48.

\bibitem{cairney}
\textbf{Cairney JWG. 1992.} Translocation of solutes in ectomycorrhizal and saprotrophic rhizomorphs.  \emph{Mycol Res}. \textbf{96}: 135-141.

\bibitem{clipson}
\textbf{Clipson N, Cairney J, Jennings D. 1987.} Phosphate uptake by cords and mycelium in the laboratory and the field.  \emph{New Phytol}. \textbf{105}: 449-457.

\bibitem{connolly}
\textbf{Connolly J, Jellison J. 1997.} Two-way translocation of cations by the brown rot fungus \emph{Gloeophyllum trabeum}.  \emph{Int Biodeterioration Biodegrad}. \textbf{39}: 181-188.



\bibitem{davidson2}
\textbf{Davidson, F. 2007.} Mathematical modelling of mycelia: a question of scale. \emph{Fungal Biology Reviews}. \textbf{21}: 30-41.

\bibitem{dreyer}
\textbf{Dreyer O, Puzio R. 2001.} Allometric scaling in animals and plants. \emph{J Math Biol}.  \textbf{43}: 144-156.

\bibitem{eamus}
\textbf{Eamus D, Jennings D. 1984.} Determination of water, solute and turgor potentials of mycelium of various basidiomycete fungi causing wood decay. \emph{J Exp Bot}. \textbf{35}: 1782-1786.

\bibitem{eamus2}
\textbf{Eamus D, Thompson W, Cairney J, Jennings D. 1985.} Internal structure and hydraulic conductivity of basidiomycete translocating organs. \emph{J Exp Bot}. \textbf{36}: 1110-1116.


\bibitem{falconer1}
\textbf{Falconer RE, Brown JL, White NA, Crawford, JW. 2005.} Biomass recycling and the origin of phenotype in fungal mycelia. \emph{Proc Roy Soc B}. \textbf{272}: 1727-1734.

\bibitem{falconer2}
\textbf{Falconer RE, Brown JL, White NA, Crawford, JW. 2007.} Biomass recycling: a key to efficient foraging by fungal colonies. \emph{Oikos}. \textbf{116}: 1558-1568.

\bibitem{fricker}
\textbf{Fricker M, Boddy L, Bebber D. 2007.} Network organisation of mycelial fungi. \emph{The Mycota VIII: Biology of the Fungal Cell}. Springer-Verlag. Pp 309-330.

\bibitem{fricker2}
\textbf{Fricker M, Lee J, Bebber D, Tlalka M, Hynes J. 2008.} Imaging complex nutrient dynamics in mycelial networks. \emph{J Microsc}. \textbf{231}: 317-331.

\bibitem{fushimi}
\textbf{Fushimi K, Verkman AS. 1991.} Low viscosity in the aqueous domain of cell cytoplasm measured by picosecond polarization microßuorimetry. \emph{J Cell Biol}. \textbf{112}: 719-725.

\bibitem{gooday}
\textbf{Gooday, GW. 1995.} The dynamics of hyphal growth. \emph{Mycol Res}. \textbf{99}: 385-394.

\bibitem{gow}
\textbf{Gow NAR, Gadd GM (Eds.) 1995.} \emph{The Growing Fungus}. Chapman and Hall, London.

\bibitem{grimmett}
\textbf{Grimmett GR, Kesten H, 1984.} Random electrical networks on complete graphs. \emph{J Lond Math Soc}. \textbf{30}: 171-192.

\bibitem{heath}
\textbf{Heath I, Steinberg G. 1999.} Mechanisms of hyphal tip growth: Tube dwelling amebae revisited. \emph{Fung Genet Biol}. \textbf{28}: 79-93.

\bibitem{howard}
\textbf{Howard, R. 1981.} Ultrastructural analysis of hyphal tip cell growth in fungi: Spitzenk\"{o}rper, cytoskeleton and endomembranes after freeze-substitution. \emph{Cell Sci}. \textbf{48}: 89-103.

\bibitem{jarrett}
\textbf{Jarrett T, Ashton D, Fricker M, Johnson N. 2006.} Interplay between function and structure in complex networks. \emph{Phys Rev Lett E}. \textbf{74}: 026116.1-026116.8.

\bibitem{jennings}
\textbf{Jennings. 1987.} Translocation of solutes in fungi. \emph{Biol Revs}. \textbf{62}: 215-243.

\bibitem{kamiya}
\textbf{Kamiya A, Bukhari R, Togawa T. 1984.} Adaptive regulation of wall shear stress optimizing vascular tree function. \emph{Bull Math Biol}. \textbf{46}: 127-137.

\bibitem{lew}
\textbf{Lew R, Levina N, Walker S, Garrilll A. 2004.} Turgor regulation in hyphal organisms. \emph{Fung Genet Biol}. \textbf{41}: 1007-1015.


\bibitem{lew2}
\textbf{Lew, R. 2005.} Mass flow and pressure-driven hyphal extension in \textit{Neurospora crassa}. \emph{Microbiology}. \textbf{151}: 2685-2692.

\bibitem{lindahl}
\textbf{Lindahl B, Finlay R, Olsson S. 2001.} Simultaneous, bidirectional translocation of $^{32}$P and $^{33}$P between wood blocks connected by mycelial cords of \textit{Hypholoma fasciculare}. \emph{New Phytol}. \textbf{150}: 189-194.

\bibitem{lopez}
\textbf{L\'{o}pez E, Buldyrev SV, Havlin S, Stanley HE. 2005.} Anomalous Transport in Scale-Free Networks. \emph{Phys Rev Lett}. \textbf{94}: 248701.1-248701.4.




\bibitem{money3}
\textbf{Money N. 1997.} Wishful Thinking of Turgor Revisited: The Mechanics of Fungal Growth. \emph{Fung Genet Biol}. \textbf{21}: 173-187.


\bibitem{money5}
\textbf{Money N. 2008.} Insights on the mechanics of hyphal growth. \emph{Fungal Biology Reviews}. \textbf{22}: 71-76.

\bibitem{nelson}
\textbf{Nelson P. 2003.} \emph{Biological Physics: Energy, Information, Life}. W.H. Freeman and Company. Pp 245-259.

\bibitem{nobel}
\textbf{Nobel PS. 1991.} \emph{Physicochemical and Environmental Plant Physiology}. San Diego: Academic Press. Pp 473-519.

\bibitem{olsson}
\textbf{Olsson S, Gray SN. 1998.} Patterns and dynamics of $^{32}$P-phosphate and labelled 2-aminoisobutyric acid ($^{14}$C-AIB) translocation in intact basidiomycete mycelia.
\emph{FEMS Microbiol Ecol}. \textbf{26}: 109-120.

\bibitem{olsson2}
\textbf{Olsson S. 2001.} Colonial growth of fungi. In \emph{The Mycota VIII: Biology of the Fungal Cell}. Eds R. Howard and N. Gow. Springer-Verlag, Heidelberg. Pp 126-141.

\bibitem{ray}
\textbf{Ray P, Green P, Cleland R. 1972.} Role of turgor in plant cell growth. \emph{Nature}. \textbf{239}: 163-164.

\bibitem{rayner2}
\textbf{Rayner ADM, Watkins ZR, Beeching JR. 1991.} Self-integration - an emerging concept from the fungal mycelium. In \emph{The Fungal Colony}. Eds. Gow, N.A.R., Robson, G.D. and Gadd, G.M. Cambridge University Press. Pp 1-24.

\bibitem{rodbard}
\textbf{Rodbard, S. 1975.} Vascular caliber. \emph{Cardiology}. \textbf{60}: 4-49.

\bibitem{savage}
\textbf{Savage, V. M., Deeds, E. J., Fontana, W. 2008.} Sizing up allometric scaling theory. \emph{PLoS Comput Biol}. \textbf{4}(9): e1000171. 


\bibitem{sherman}
\textbf{Sherman TF. 1981.} On connecting large vessels to small: the meaning of Murray's Law. \emph{J Gen Physiol}. \textbf{78}: 431-453.

\bibitem{steinberg}
\textbf{Steinberg G. 2006.} Hyphal Growth: a Tale of Motors, Lipids, and the Spitzenk\"{o}rper. \emph{Eukaryotic Cell}. \textbf{6}: 351-360.

\bibitem{thompson}
\textbf{Thompson W, Brownlee C, Jennings D, Mortimer A. 1987.} Localized, Cold-Induced Inhibition of Translocation in Mycelia and Strands of \textit{Serpula lacrimans}. \emph{J Exp Bot}. \textbf{38}: 889-899.

\bibitem{thompson2}
\textbf{Thompson W, Eamus D, Jennings DH. 1985.} Water flow through the mycelium of \emph{Serpula lacrimans}. \emph{Trans Brit Mycol Soc}. \textbf{84}: 601-608.

\bibitem{tlalka}
\textbf{Tlalka M, Watkinson SC, Darrah PR, Fricker MD. 2002.} Continuous imaging of amino-acid translocation in intact mycelia of \textit{Phanerochaete velutina} reveals rapid, pulsatile ßuxes. \emph{New Phytol}. \textbf{153}: 173-184.

\bibitem{tlalka2}
\textbf{Tlalka M, Hensman D, Darrah PR, Watkinson SC, Fricker MD. 2003.} Noncircadian oscillations in amino acid transport have complementary profiles in assimilatory and foraging hyphae of \textit{Phanerochaete velutina}. \emph{New Phytol}. \textbf{158}: 325-335.

\bibitem{tlalka3}
\textbf{Tlalka M, Bebber DP, Darrah PR, Watkinson SC, Fricker MD. 2008.} Quantifying dynamic resource allocation illuminates foraging strategy in \textit{Phanerochaete velutina}. \emph{Fung Genet Biol}. \textbf{45}: 1111-1121.


\bibitem{wells}
\textbf{Wells J, Boddy L. 1995a.} Effect of temperature on wood decay and translocation of soil-derived phosphorus in mycelial cord systems. \emph{New Phytol}. \textbf{129}: 289-297.

\bibitem{wells2}
\textbf{Wells J, Boddy L, Evans R. 1995b.} Carbon translocation in mycelial cord systems of \textit{Phanerochaete velutina}. \emph{New Phytol}. \textbf{129}: 467-476.

\end{thebibliography}
\end{document}